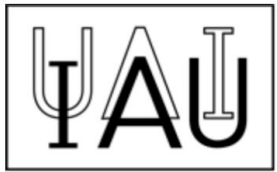 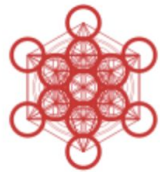 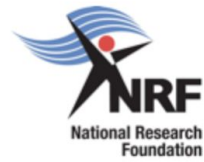

# International Coordination of Multi-Messenger Transient Observations in the 2020s and Beyond
Kavli-IAU White Paper

S. Bradley Cenko[1] (co-chair), Patricia A. Whitelock[2] (co-chair), Laura Cadonati[3], Valerie Connaughton[4], Roger Davies[5], Rob Fender[5], Paul J. Groot[6], Mansi M. Kasliwal[7], Tara Murphy[8], Samaya Nissanke[9], Alberto Sesana[10], Shigeru Yoshida[11] and Binbin Zhang[12]

[1] Astrophysics Science Division, NASA Goddard Space Flight Center, Mail Code 661, Greenbelt, MD 20771, USA; Joint Space-Science Institute, University of Maryland, College Park, MD 20742, USA
[2] South African Astronomical Observatory, P.O. Box 9, Observatory, 7935 Cape Town, South Africa; Department of Astronomy, University of Cape Town, 7701 Rondebosch, South Africa
[3] Georgia Institute of Technology, USA
[4] SMD - Astrophysics Division, NASA HQ, 300 E St SW, Washington, DC 20546, USA
[5] Department of Physics, University of Oxford, Denys Wilkinson Building, Keble Rd., Oxford, OX1 3RH, UK
[6] Department of Astrophysics, IMAPP, Radboud University Nijmegen, PO Box 9010, 6500 GL Nijmegen, the Netherlands; Department of Astronomy and Inter-University Institute for Data Intensive Astronomy, University of Cape Town, 7701 Rondebosch, South Africa; South African Astronomical Observatory, P.O. Box 9, Observatory, 7935 Cape Town, South Africa
[7] Division of Physics, Mathematics and Astronomy, California Institute of Technology, Pasadena, CA 91125, USA
[8] Sydney Institute for Astronomy, School of Physics, University of Sydney, NSW, 2006, Australia
[9] GRAPPA, Anton Pannekoek Institute for Astronomy and Institute of High-Energy Physics, University of Amsterdam, Science Park 904, 1098 XH Amsterdam & Nikhef, Science Park 105, 1098 XG Amsterdam, The Netherlands
[10] Department of Physics G. Occhialini, University of Milano - Bicocca, Piazza della Scienza 3, 20126 Milano, Italy
[11] Department of Physics, Graduate School of Science and International Center for Hadron Astrophysics Chiba University, Japan
[12] School of Astronomy and Space Science Nanjing University, China

## Summary

This White Paper summarizes the discussions from a five-day workshop, involving 50 people from 18 countries, held in Cape Town, South Africa in February 2020. Convened by the International Astronomical Union's Executive Committee Working Group on Global Coordination of Ground and Space Astrophysics and sponsored by the Kavli Foundation, we discussed existing and potential bottlenecks for transient and multi-messenger astronomy, identifying eight broad areas of concern. Some of these are very similar to the challenges faced by many astronomers engaging in international collaboration, for example, data access policies, funding, theoretical and computational resources and workforce equity. Others, including, alerts, telescope coordination and target-of-opportunity implementation, are strongly linked to the time domain and are particularly challenging as we respond to transients. To address these bottlenecks we offer thirty-five specific recommendations, some of which are simply starting points and require development. These recommendations are not only aimed at collaborative groups and individuals, but also at the various organizations who are essential to making transient collaborations efficient and effective: including the International Astronomical Union, observatories, projects, scientific journals and funding agencies. We hope those involved in transient research will find them constructive and use them to develop collaborations with greater impact and more inclusive teams.

**Table of Contents**

## Motivation and Objectives

With the recent discoveries of an electromagnetic counterpart to the binary neutron star merger gravitational-wave source GW170817 ([Abbott et al. 2017](#)), as well as the association between a flaring high-energy blazar TXS 0506+056 and a high-energy neutrino event ([Aartsen at al. 2018](#)), a new era of multi-messenger astrophysics is upon us. Observations of multi-messenger and transient sources in the coming decade promise to address some of the most pressing questions in (astro)physics, including testing General Relativity, the origin of the heavy elements, the equation of state of dense matter, and the source(s) of ultra-high energy cosmic rays.

Fully capitalizing on the opportunities offered as new facilities come online (e.g., Vera C. Rubin Observatory, the Extremely Large Telescopes and the Square Kilometre Array) and current facilities are upgraded (e.g., Advanced LIGO+, Advanced Virgo+, and the planned IceCube-Gen2) will require international coordination at a scale not currently practiced. Recognizing this, the IAU Executive Committee Working Group on Coordination of Ground and Space Astrophysics, with financial support from the Kavli Foundation, organized a workshop at the South African Astronomical Observatory (SAAO) in Cape Town, South Africa, from 2 to 7 February, 2020.

Fifty representatives of the major observatories, projects, and scientific interests [see Appendix 3 for a list of the  projects and observatories represented and Appendix 4 for the full list of Participants] involved in transient and multi-messenger astronomy gathered to discuss how to take this exciting and rapidly developing field forward in a strategic and inclusive manner. To capture the results of this workshop, we have produced this White Paper that includes both recommendations and best practices for the community. It is not intended to be the final word on this complex subject, but instead as one contribution to an evolving conversation about how to optimize the scientific return from this exciting new era in astronomy.

Most of the recommendations below have been associated with one or more IAU Division, Commision and/or Working Group, which we ask to consider whether drafting an IAU Resolution, involving one or more of the recommendations, is desirable. If it is, they should, in liaison with other stakeholders, draft the appropriate wording, for consideration by the next IAU General Assembly (to be held in Busan in the Republic of Korea in August 2021). Note that (assuming the resolution has no budgetary implications for the IAU) these drafts should be submitted to the General Secretary six months before the General Assembly.

**Bottleneck Overview**

Prior to the workshop, the Scientific Organizing Committee [authors of this White Paper] met and held a number of sessions to structure the workshop. In addition to solicited talks on the lessons learned from the major projects working in transient/multi-messenger astronomy and on the scientific opportunities in the coming decade, the SOC identified a series of "bottlenecks" that are currently limiting scientific progress in these areas [see Appendix 1 for the full Workshop Program]. While there was clear overlap between topics and common themes that emerged throughout, we have structured this document following the bottleneck areas discussed in extended sessions at the workshop.

**B1 - Transient Alerts and Communication**

Timely notification of newly discovered transients (or, for that matter, interesting behavior of known sources) is critical to our understanding of these events. Well-designed systems enable an efficient marshaling of observational resources larger than could be provided by any single individual or collaboration, resulting in greatly improved scientific understanding. Much like peer-reviewed publications, effective transient notification systems also provide motivation to observers by assigning scientific "credit" to those enabling discovery by others (e.g., precisely localizing gamma-ray burst afterglows, which can then subsequently be observed with sensitive, narrow-field facilities).

The most well-known example was the development 28 years ago of the Gamma-Ray Coordinates Network (GCN). Born of necessity due to the failure of an on-board recording device on the Compton Gamma-Ray Observatory, the GCN system currently provides prompt notifications of transient discoveries from 13 distinct "streams", including both high-energy satellites (e.g., *Swift, Fermi, INTEGRAL*) and ground-based observatories (Advanced LIGO-Virgo, IceCube, HAWC). In addition to these machine-readable Notices, human-written Circulars describing observing results are published by the community at a rate of several hundred per month.

A more recent highly successful example is the Transient Name Server (TNS), which on 1 January 2016 became the official IAU mechanism for reporting astronomical transients, and in particular designating names for supernovae. TNS can receive both manual (e.g., from amateurs, who still discover a number of nearby supernovae each year) and fully automated (e.g., from large professional surveys such as ATLAS, Pan-STARRS, and ZTF) transient discoveries, automatically associating observations of objects imaged by different surveys. Upon spectroscopic classification and confirmation of a supernova origin (as opposed to, say, a previously unknown flare star), sources are provided an official supernova designation (e.g., SN2020A, SN2020B, …). Currently approximately 3000 optical transients are reported each month to TNS, compared to less than 500 total[13] reported to the Central Bureau for Astronomical Telegrams for the entire year of 2011.

Impending event rate increases in transient and multi-messenger discoveries outside the (current) purview of TNS require new and/or updated systems with similar capabilities: name service, machine-readable communication, etc. Given the widely varying needs of different communities within transient/multi-messenger astronomy, a single "one-size fits all" solution is unlikely to be optimal. We therefore make the following recommendations:

*Alerts/Communication Recommendation 1:* Despite its role as a pioneer in automated event notification, there is a clear need for upgrades to the GCN system to accommodate the new multi-messenger era. This includes (but is not limited to): an automated name server to correlate events found by different facilities and/or at different wavelengths; increased machine-readability for GCN Circulars (e.g., standardize formats, tagging specific events, flux measurements / upper limits); and a means to query results based on source, position, and/or time. A number of ongoing efforts include GCN upgrades/augmentation as one (of many) goals, including the Time-Domain Astronomy Coordinate Hub (TACH) and the Scalable Cyberinfrastructure for Multi-Messenger Astrophysics (SCiMMA). Coordination amongst these efforts is critical going forward; the IAU should consider the formal adoption

---

[13] Including both spectroscopically confirmed supernovae and unconfirmed optical transients.

of a naming convention for such sources and endorsing a centralized clearing house similar to TNS. [DivD WG:SNe DivB]

*Alerts/Communication Recommendation 2*: In the near future, similar notification capabilities will be necessary for X-ray (e.g., SRG/eROSITA, Einstein Probe) and radio (e.g., MeerKAT, ASKAP, SKA) transients. Under the auspices of major projects in these areas, we recommend that the relevant major projects work with their communities to develop desired approaches, utilizing lessons learned from e.g., the optical and gamma-ray communities. An example is an effort in the fast radio burst community, where public alerts (utilizing IVOA standards, Petroff et al. 2017) will be incorporated into TNS. These proposals can then be brought to the IAU [DivB Com B2, B4; DivD ] for official endorsement.

*Alerts/Communication Recommendation 3:* Standards are critical to ensuring that heterogeneous nodes in the transient/multi-messenger ecosystem can communicate efficiently and effectively. We strongly recommend that the transient/multi-messenger community work with the International Virtual Observatory Alliance (IVOA) to build its standards. IVOA standards should follow the procedure described in the IVOA Document Standards (Genova et al. 2017), with at least two reference implementations and available validation tools. In this process it is critical to have involvement and support of the community, which can be accomplished in a variety of ways (technical/scientific interchanges, training schools, etc.). [DivB ComB2, DivD ComD1 WG:SN]

*Alerts/Communication Recommendation 4:* The ultimate arbiter of scientific analysis is the peer review process. As such, scientific journals have a unique opportunity to play a critical role in ensuring that published data is broadly accessible. We strongly recommend the IAU work with relevant journals to establish uniform standards for data reporting on transient and multi-messenger sources (as well as static sources, for that matter). This will facilitate large astronomical databases (CDS, NED, etc.) in being able to ingest such measurements, making them broadly available to astronomers across the world. [DivB ComB2, WG:Information Professionals].

## B2 - Data Policies

While an issue for the astronomy community as a whole, data availability and accessibility is particularly important in the transient/multi-messenger world, where time-critical observations cannot be repeated. Widely available and accessible data products greatly enhance the scientific return of any project. Furthermore, in a [recent survey](#) of those utilizing NASA missions for multi-messenger science, a strong preference was expressed for "open" data policies (corresponding to proprietary periods less than or equal to 1 month). Lowering barriers to data access is particularly critical for astronomers from developing countries ([Peek at al. 2019](#)).

At the same time, multiple complex factors often compete with the desire for broadly "open" data policies, in some cases for good reasons. All projects have a legitimate interest in ensuring that those responsible for funding a facility reap the benefit of its scientific return, whether this be individual scientists, institutions, or even taxpayers of a given nation. There also can be a cultural disconnect between the high-energy physics community and the astronomy community, possibly in part due to the distinction sometimes found in astronomy between those operating a facility and those performing scientific analysis (e.g., HST). To balance these competing interests, we recommend the following:

*Data Policies Recommendation 1:* we recommend the IAU build on their [Resolution on Open Access](#), made in Sydney in 2003, making specific reference to the importance of limited data proprietary periods for the effective follow-up of transients and other time-critical observations. [DivB ComB2]

*Data Policies Recommendation 2:* Towards this end, data availability is not sufficient: data should be FAIR (findable, accessible, interoperable, and reusable). Good archives and supporting software are thus essential. As a result, even data policies with modest proprietary periods can contribute to the broad goals of open data. For example, by releasing not only raw data but high-level data products in a queryable database, the Sloan Digital Sky Survey has distinguished itself as one of the most scientifically productive projects in astronomy in the last several decades. Organizations such as ESCAPE, working with all the major astronomy infrastructures in a region, will have an increasingly important role to play.  [DivB ComB2]

*Data Policies Recommendation 3:* Large projects in the USA and Europe should pay particular care towards supporting open data policies. Other countries including China and India are rapidly developing observational capabilities in astronomy, some are already using current and next generation flagship facilities in the USA and Europe as models for their data policies. Long proprietary periods based on country of origin, e.g. SKA,  and/or specific subscription models, such as is planned for Rubin Observatory, and a lack of funding for worldwide data distribution will set a deleterious precedent  at this critical time. Open-access to data associated with follow-up observations is  just as important, but it is the major projects that set the tone and expectation of others.  [DivB ECWG:Global Coordination]

## B3 - Follow-Up Spectroscopy

At optical wavelengths (which currently dominate transient discoveries), discovery is largely done via broadband imaging, with no, or extremely limited, information about the source spectrum. Yet with some notable exceptions (e.g., cosmology with photometric supernovae), most scientific analyses require optical (or ultraviolet/near-infrared) spectra in order to obtain any physical understanding of their nature. From redshift measurement and classification to ejecta composition and velocity, the power of spectra is readily apparent to all practicing astronomers.

Yet even with the current generations of discovery facilities, the community's capacity to discover new sources is greatly outpacing our ability to understand them - even for relatively bright supernovae (V< 19 mag), only ~ 10% obtain a spectroscopic classification (Kulkarni 2020). This situation will only be exacerbated as more sensitive facilities like Rubin Observatory come online in the coming years. While this issue is not as acute for other wavelengths (e.g., X-rays and radio), that also may change in the not too distant future (e.g., SKA).

Lack of access to (in particular) optical spectroscopy is a clear bottleneck towards the scientific understanding of transients and multi-messenger sources. To address this, we recommend the following:

*Follow-up Spectroscopy Recommendation 1:* A concerted effort be made around the globe, enlisting philanthropic and public funding opportunities, to increase the capacity for spectroscopic follow-up. For current generation facilities, this does not require particularly large telescopes: currently the 1.5m Palomar telescope equipped with a very low-resolution (R~100) integral-field spectrograph (SED Machine) leads the world in supernova classification. Dedicated instruments on modest-aperture facilities can have a very large impact here, particularly if economies of scale could be harvested (e.g., instrument copies on multiple telescopes). [DivB ECWG:Global Coordination]

*Follow-up Spectroscopy Recommendation 2:* In a similar vein, lack of spectroscopic follow-up offers an opportunity for developing countries to make major contributions to transient and multi-messenger science. Our experience has shown that integration into a larger network of observational facilities has proved a successful strategy for long-term engagement (e.g., the GRANDMA and GROWTH collaborations). Major investments in the development of "Observatory-level" capabilities, as is being done in South Africa, is also an effective strategy for some nations (though may be less feasible for many). [ECWG:Global Coordination]

*Follow-up Spectroscopy Recommendation 3:* Despite lacking sufficient global capacity in this area, there is still some degree of inefficiency in that duplicative observations of the same source may be obtained. While this is inevitable and in some cases desirable (e.g., to cross-check results), improved communication protocols (see *Alerts/Communication* and *Global Coordination Recommendations*) would nonetheless result in a more efficient usage of limited resources. [DivB ECWG:Global Coordination]

*Follow-up Spectroscopy Recommendation 4:* The limiting magnitude of current optical surveys such as ASAS-SN, Pan-STARRS, ATLAS, and ZTF is well-matched to obtaining classification spectra with moderate-aperture telescopes. While it is still possible, we recommend astronomers undertake (and facilities support) spectroscopic measurements on large, unbiased samples (e.g., well-defined flux- and/or magnitude-limited samples), which can be used to train machine-learning models, to improve our understanding of transient demographics and improve the performance of machine-learning classifiers. Such results will be critical to guide the usage of more limited follow-up with large-aperture facilities in the era of Rubin Observatory - hence the time critical nature of such endeavors. [DivB ECWG:Global Coordination]

**B4 - Telescope Coordination**

Transient and multi-messenger science inherently requires complex observational data, spanning a diverse range of wavelengths, time scales, and (of course) cosmic messengers. Combined with the ephemeral nature of most sources, this presents the community with a unique set of challenges in acquiring datasets that span across observing technique, funding agency, ground vs. space, etc. The success of recent visualization packages such as the Treasure Map suggest that such a need remains to be filled by the community.

To simplify the acquisition of the observational datasets necessary to make ground-breaking progress in these areas, we make the following recommendations:

*Telescope Coordination Recommendation 1:* We recommend the IAU endorse a common format for all observatories to report previous and planned observations, namely the standard developed by the IVOA (**ObsLocTAP**, Salgado et al. 2020). A number of space-based facilities have begun to implement this protocol already (e.g., *Chandra, NuSTAR*); broad-based buy-in from current and future facilities would make the protocol even more useful to the broader transient and multi-messenger community. When existing standards are found limiting or inadequate, collaborative efforts should be made to improve or replace them, perhaps through the IVOA, to ensure interoperability. [DivB, ComB2]

*Telescope Coordination Recommendation 2:* Multi-observatory datasets can be particularly difficult to acquire, as proposals face not only the issue of "double jeopardy" but also the logistical challenge of coordinating different dates for each observatory (e.g., semester vs. year cycles, offset cycle starting dates, etc.). A number of facilities offer joint proposal opportunities to mitigate this issue (e.g., *Chandra+HST*, *VLA+XMM,* etc.). We strongly recommend that all projects, particularly large facilities, pursue joint proposal opportunities aggressively and early in their life cycle. As an example, in the USA, the Las Cumbres Observatory is partnering with NOIRLab facilities (Gemini Observatory, SOAR Observatory, and the CSDC) to form AEON, the Astronomical Event Observatory Network.  AEON will provide rapid, flexible, programmable access to multiple telescope follow-up facilities. The AEON partners are working to form a single time allocation process which could be expanded to include other facilities. For timely follow-up, intercontinental collaboration would be especially beneficial, e.g. between the telescopes in South Africa and/or Australia and various facilities in Chile. [DivB ECWG:Global Coordination]

## B5 - International Funding and Collaboration

The scope of the scientific questions addressed currently by the transient and multi-messenger communities is remarkably broad - ranging from fundamental physics (the expansion history of the Universe; the equation of state of dense matter) to some of the most long-standing questions in astronomy (the origin of cosmic rays, heavy elements, and relativistic jets). It is not surprising, then, that the facilities required to address many of these questions are inherently large, often requiring multi-national collaborations to support them.

Furthermore, given the rapid evolution of many transient and multi-messenger sources of interest (in particular those on sub-day time scales), observing facilities distributed around the globe are essential. Thus our community is more dependent on the ease and availability of international collaborations than many other astronomical disciplines.

Spinning up new collaborations, ranging from two individual researchers to the largest multinational observing facilities, across international boundaries presents a unique set of challenges. From language and cultural barriers to funding and legal issues, difficulties abound. There are often not easy solutions to these issues; however, we recommend the following as positive steps towards enabling international collaboration in this area:

*International Collaboration Recommendation 1:* We recommend the IAU endorse and support international networks of people with common science goals to exploit existing facilities and/or propose new ones, facilitated for example through their Working Group on Coordination of Ground and Space Astrophysics. With improved capabilities for remote collaboration, such networks can be particularly powerful in transient studies by capitalizing on time-zone differences (e.g., responding to time-critical alerts during local day time). Often funding agencies are more receptive to funding people rather than hardware (particularly when that hardware funding may be spent abroad). These are also natural projects to develop capacity in areas lacking such infrastructure (e.g., African VLBI). International efforts such as the PIRE program and the Newton Fund should be strongly encouraged to continue (and/or be expanded). [ECWG:Global Coordination]

*International Collaboration Recommendation 2:* Training early-career scientists and engineers is critical towards a self-sustaining global community of astronomy. The IAU International Schools for Young Astronomers (ISYA) have been highly successful in this area; we encourage the IAU to work in partnership with other organizations and facilities (e.g., ESO and NASA) so as to increase cross-fertilization across the major communities, but also to facilitate the involvement of young people from less developed nations. [DivC OYA]

*International Collaboration Recommendation 3:* An additional major challenge towards scientific advancement is the difficulty in establishing large, international projects. Even once such efforts have passed the proposal stage, creating an appropriate legal entity with an acceptable management structure can often delay projects more than hardware difficulties. There are several intergovernmental treaty level organisations that can play this role, but not all countries are prepared to subscribe to such organisations. There are a number of other models, besides treaty level organisations, that have been shown to work e.g. ERICs for CTA, umbrella entities such as AUI or AURA that join with a treaty level organisation (ESO), e.g. ALMA (but note that SKA needed to create its own IGO). We therefore recommend that the IAU commission a separate group to study this issue and generate a report on best practices for scientists and funding agencies, as the full scope of this issue was beyond what could be addressed at our workshop. [ECWG:Global Coordination]

*International Collaboration Recommendation 4:* We encourage scientists to pursue alternatives to standard (i.e., governmental) funding routes to enable future projects of modest scale. Philanthropic funders are playing an increasing role in supporting scientific endeavors, and often do not suffer from the same limitations on international cooperation as governmental organizations. Similarly, in some cases corporations may be valuable partners for projects, as data generated by our facilities is in some cases of interest beyond the astronomical community (e.g., BlackGEM data is being archived by Google for no charge, due to the perceived educational and public relations value of the BlackGEM dataset). [not suitable for IAU Resolution]

*International Collaboration Recommendation 5:* Satellite access to low-Earth orbit (LEO) has decreased in cost dramatically in recent years, enabling novel possibilities for transient astronomy (e.g., CubeSat constellations). However, increased LEO usage, particularly by industry, also presents challenges: streaked images for ground-based telescopes, radio frequency interference, increased risk of collision, etc. We commend the IAU for taking a proactive approach to [address](#) satellite constellations. Continued IAU engagement with all stakeholders (industry, science, and military) will be necessary to ensure all parties continue to function as stewards of both LEO and the night sky, in addressing these important challenges. In particular, to minimize the likelihood of unforeseen satellite collisions, we encourage the IAU to explore the possibility of a separate group to coordinate civil information gathering and dissemination of satellite orbit data. [DivBC ComB7, Executive Committee]

## B6 - Target-of-Opportunity Implementation

For many (though not all) science topics in transient and multi-messenger astronomy, rapid response observations are required. Examples range from the most distant gamma-ray burst afterglows from the epoch of reionization, to newly discovered comets in our solar system. Such observations are by definition not possible to schedule in advance, necessitating the implementation of "target-of-opportunity" (ToO) interrupt schemes.

Given that each observing facility presents a unique set of challenges in implementing a ToO program, no "one-size-fits-all" solution will be applicable for all projects. Nonetheless, based on our collective experience in the field we highlight the properties we find in common amongst the most successful ToO programs at facilities across the globe. Rather than a series of recommendations, we consider this a list of best practices for facilities to draw from when considering their unique circumstances in implementing transient/multi-messenger science:

*ToO Implementation Best Practice 1:* For new / planned facilities, ToO programs function most effectively when treated as part of the requirements definition process in the early stages of formulation. Attempts to implement ToO programs in existing facilities inevitably encounter fundamental limits in e.g., response time, data latency, etc. that may well have been circumvented with little to no change in project cost, if addressed at an earlier stage. [DivB]

*ToO Implementation Best Practice 2:* The broad properties of the ToO program, including the total amount of time available, the allowed response time, etc., should be driven by the science (and not, e.g., programmatic considerations) whenever feasible. Thus, "artificial" caps (both upper and lower limits) on ToO time are not placed on observers as part of a proposal call (even if ultimately they are not strictly adhered to). [DivB]

*ToO Implementation Best Practice 3:* The time allocation policy for ToO observations (both through regular peer-reviewed proposals and via Director's Discretionary opportunities) is transparent, to ensure that everyone is aware of how choices are made. [DivB]

*ToO Implementation Best Practice 4:* Proprietary periods are limited so that data may be used by as broad a section of the community as possible (see *Data Policies Recommendation 1)* [DivB]

*ToO Implementation Best Practice 5:* ToO policies are constructed to encourage groups to pool resources, work together (where sensible), and maximize the science possible from any given dataset. For example, the capability to propose as "Co-PIs" with *HST* allows early career scientists, who might otherwise struggle to compete, to be able to collaborate, while still demonstrating scientific leadership (e.g., for promotion or in tenure packets). [DivB]

*ToO Implementation Best Practice 6:* Data products are available promptly, and software tools are in place to rapidly produce publication-quality results (see also *Data Policies Recommendation 2).* [DivB]

## B7 - Theoretical and Computational Resources

Theoretical advances have always been critical in the development of transient and multi-messenger astronomy. For example, the development in the last several decades of catalogs of waveforms generated by numerical simulations of merging compact objects is a fundamental ingredient of the most sensitive template-based searches employed by ground-based gravitational wave detectors. More recently, both analytic and numerical calculations of the predicted electromagnetic emission from kilonovae - the transient event powered by r-process nucleosynthesis accompanying the merger of two neutron stars - played a key role in guiding the observational campaigns that discovered and characterized the UV/optical/NIR counterpart to GW170817.

Theory will continue to play a key role in multi-messenger astronomy as we look forward to the launch of the Laser Interferometer Space Antenna (LISA) mission in approximately 10 years. Much like high-frequency GW detectors, refinements in numerical simulations of merging supermassive black holes will be necessary to identify electromagnetic counterparts in the large localizations error regions.

Towards supporting a robust and sustainable theory program in transient and multi-messenger astrophysics, we make the following recommendations:

*Theory/Computation Recommendation 1:* A robust investment in theoretical research, ideally some significant fraction of that provided to observational facilities (e.g., Kollmeier et al. 2020), is necessary for the long-term success of transient and multi-messenger astrophysics. We encourage the IAU to endorse such an investment in theory from relevant funding agencies.  [DivBDGHJ - but much broader than transients]

*Theory/Computation Recommendation 2:* Modelling of binary mergers, in particular the supermassive black holes relevant for the LISA mission, is extremely complex, involving a wide range of time and size scales and a diverse set of underlying (astro)physics. As a result, state-of-the-art simulations require increasingly large computational allocations. The current PI-funded model is not well-suited to such a problem, where major resources are necessary from many institutions. We encourage the IAU to endorse new models for relevant funding agencies suitable for the large scale of the problem.  [DivB Exec Comm]

*Theory/Computation Recommendation 3:* Training and workforce retention of numericists is a major issue for the future of this field. To encourage as broad and international a workforce as possible, we recommend current practitioners work to develop online courses with regular updates (given the fast-evolving nature of the field).  [DivBC OYA OAD]

*Theory/Computation Recommendation 4:* Infrastructure, in particular internet bandwidth, is a fundamental limitation towards the full harnessing of talent in developing countries. Efforts to expand internet access around the globe should be strongly encouraged. It may be possible to use new observational facilities in developing countries as leverage for such investments, as they are critical to the success of such projects. Every effort must also be made to ensure that doing so does not degrade sky access and therefore negatively impact astronomy (see also *International Collaboration Recommendation 5)*.  [DivB OAD ECWG:Global Coordination]

**B8 - Diversity, Equity, Inclusion and Workforce Development**

Even aside from moral considerations, a diverse and inclusive working environment has been repeatedly demonstrated to foster more effective teams (Hunt at al. 2015). To tackle the major questions posed in the transient / multi-messenger worlds, developing and then tapping into the skill set of the broad population of historically under-represented groups will be required. However, there is growing evidence of bias, detrimentally affecting members of these groups, in the allocation of resources, including telescope time (e.g. Reid & Strolger 2019, Johnson & Kirk 2020).

Transient and multi-messenger science poses a unique set of challenges in this area. While the astronomy community is still on a long path towards broadly establishing more equitable and inclusive collaborations, the high-energy physics community, where many key multi-messenger facilities have been historically located, lags behind even this standard. The international nature of many collaborations, stretching across diverse cultural norms, makes these issues particularly acute.

Nonetheless, we see reason for some optimism. In the United States, the Decadal Survey process has for the first time solicited white papers in the area of the "State of the Profession and Societal Impacts", with a goal to assess the health of the community, including topics such as demographics, diversity and inclusion, workplace climate, workforce development, education, and public engagement. New flagship facilities, such as Rubin Observatory, have invested significantly in both creating a diverse and inclusive collaboration.  Countries such as South Africa that have invested heavily in astronomy and transformation and have seen this investment pay off, both in terms of world-class facilities, but more importantly in training a diverse set of citizens with technical skills that translate beyond the astronomy world.

Entire workshops could be (and have been!) devoted to the topics above. Here we highlight some of the recommendations arrived at during our week-long workshop:

*DEI Recommendation 1:* Sharing of best practices is critical. Even well-intentioned leadership may be unaware of what needs to be done (and why). The Multi-Messenger Diversity Network (MDN) was funded in the USA by the National Science Foundation to perform this task for USA-based collaborations, and has been quite successful in this endeavor. An international analog, formed with the support of (or under the auspices of) the IAU would play an important (and missing) role for international collaborations. [Exec Comm on Astronomy for Equity and Inclusion, Exec Comm on Women in Astronomy]

*DEI Recommendation 2:* In a similar vein, the IAU may serve as a repository for best practice documents adopted by effective collaborations, with input from the [IAU Executive Working Group on Equity and Inclusion](#), as well as from the collaborations. A template code of conduct and collaboration framework agreement could be officially endorsed by the IAU, and used as a starting point by future collaborations. [Exec Comm on Astronomy for Equity and Inclusion, Exec Comm on Women in Astronomy]

*DEI Recommendation 3:* Workforce and Leadership training should be an integral part of all transient and multi-messenger collaborations, across a broad agenda. This should include skills beyond those simply necessary for day-to-day activities, as many early career scientists will ultimately pursue opportunities outside astronomy. Leadership skills are particularly important for under-represented groups, where the fractional representation in leadership positions is even lower than the overall fraction. [Exec Comm on Astronomy for Equity and Inclusion, OYA]

*DEI Recommendation 4:* In order to deal with issues of bias in the allocation of telescope time and other resources, we recommend a combination of training reviewers to deal with unconscious bias and, where practical, the use of dual-anonymization to eliminate even the possibility of bias. [Exec Comm on Astronomy for Equity and Inclusion, Exec Comm on Women in Astronomy]

*DEI Recommendation 5:* The transient and multi-messenger community (as well as the astronomy community more broadly) needs to improve recognition for contributions outside published manuscripts. Examples include instrumentation work, software development/maintenance, and also public education and outreach activities. From society prizes to hiring and promotional decisions, such vital contributions are traditionally under-valued. [DivB Exec Comm]

*DEI Recommendation 6:* We need to develop better ways of measuring the effectiveness of our methods of promoting diversity and developing improved methods. We recommend that the IAU encourage joint working with social scientists to define improved metrics in this area. [Exec Comm on Astronomy for Equity and Inclusion, Exec Comm on Women in Astronomy]

## Conclusion and General Recommendations

We hope that the outcomes of this workshop serve as a starting point towards an ongoing discussion around enabling discovery in transient and multi-messenger astronomy. While we found it to be an extremely stimulating environment, as the focus of the meeting was on discussion, we were limited in the number of attendees we could invite. Surely others have good ideas that we did not think of.

The limited attendance, together with the rapidly changing nature of our field, motivate us to issue one final recommendation:

*General Recommendation 1:* We enthusiastically encourage the IAU to continue supporting such workshops, on a regular basis (~2 years), with a rotating cast of attendees. Such a series of workshops would help reinforce that enabling international collaboration in this area is an effort that requires constant vigilance and continuous improvement on the part of the transient and multi-messenger communities. [ECWG:Global Coordination]

## Acknowledgements

We are very grateful to the Kavli Foundation for a very generous donation that made the workshop possible. Thanks to the South African Radio Astronomy Observatory (SARAO) for their sponsorship of social events, and the SAAO for their superb organization and hospitality.  In particular we are grateful to Christian Hettlage for his management of the IT associated with the meeting, to Nazli Mohamed for handling registration and most of the organizational details, to Glenda Snowball for dealing with the finance, and to Valencia Cloete for arranging the trip to Sutherland.  Thanks also to Jessymol Thomas and the following students for their help with the presentations: Miriam Nyamai, Simon de Wet, David Modiano, Danté Hewitt, Irvin Martínez and Orapeleng Mogawana.

**Appendices**

## 1      Workshop Program (Venue: SAAO Auditorium)

**Monday, 3 February**

 9:00 – 9:10    Welcome from SAAO (Patricia Whitelock), IAU (Ewine van Dishoeck) and Kavli Foundation (Christopher Martin)
 9:10 – 9:20    Aims, Objectives and Outcomes of the Workshop – Brad Cenko and Patricia Whitelock

**Session 1: Radio Transients   Chair: Tara Murphy**
 9:20 – 9:50    New developments in coherent radio transient science – Sarah Burke Spolaor
 9:50 – 10:20    Radio transient surveys: serendipitous vs. targeted – Rob Fender (via Zoom)
10:20 – 10:50  Discussion: Prospects for coherent and incoherent radio transient surveys
                Guided Discussion Leaders: Ben Stappers and Tara Murphy
10:50 – 11:20  Break

**Session 2: Alerts, Communication, and Triggering   Chair: Brad Cenko**
11:20 – 12:20  Bottlenecks I: Alerts/Communication/Triggering and active follow-up
                Guided Discussion Leaders: Tara Murphy and Patrick Woudt
12:20 – 12:35  Summary Recommendations – Brad Cenko
12:35 – 13:30  Lunch

**Session 3: High-Energy Neutrinos   Chair: Valerie Connaughton**
13:30 – 14:00  Lessons Learned: The Fermi Gamma-Ray Space Telescope – Liz Hays
14:00 – 15:30  High-Energy Neutrinos
14:00 – 14:30  Current and Future Neutrino Detectors – Ignacio Taboada
14:30 – 15:00  Science with Future Neutrino Detectors – Anna Francowiak
15:00 – 15:30  Discussion and Questions
15:30 – 16:00  Break

**Session 4: Data Protection / Rights   Chair: Valerie Connaughton**
16:00 – 17:00  Bottlenecks II: Data Protection/Rights
                Guided Discussion Leaders: Brad Cenko and Ranpal Gill
                Topics:
                The Swift Model (Brad Cenko)
                LIGO, Virgo, and Public Data Access (Johannes van den Brand)
                The Rubin Observatory (LSST) Data Rights Policy (Ranpal Gill)
                Plans for Sitian (Roberto Soria)
17:00 – 17:30  Summary Recommendations – Valerie Connaughton

**Tuesday, 4 February**

**Session 5: Stellar Explosions    Chair: Paul Groot**

9:00 –  9:20    ["Traditional" Core-Collapse and Thermonuclear Supernovae](#) – Avishay Gal-Yam
9:20 –  9:40    [The Expanding Landscape of Stellar Explosions](#) – Brad Cenko
9:40 – 10:00    [The new TeV window into Gamma-ray Bursts](#) – Ulisses Barres de Almeida
10:00 – 10:30  Discussion and Questions
10:30 – 11:00  Break

**Session 6: Follow-Up Spectroscopy    Chair: Paul Groot**

11:00 – 12:00  Bottlenecks III: Follow-Up Spectroscopy
Guided Discussion Leaders: John Blakeslee and Rob Ivison
Topics:
[Gemini in the Era of Multi-Messenger Astronomy](#) (John Blakeslee)
[ESO's Plans for Transient and Multi-Messenger Science](#) (Rob Ivison)
[SAAO/SALT's Intelligent Observatory Program](#) (Stephan Potter)
12:00 – 12:30  Summary Recommendations – Paul Groot
12:30 – 13:30  Lunch

**Session 7: Science with Optical Transient Surveys    Chair: Brad Cenko**

13:30 – 14:00  [Lessons Learned: ZTF](#) – Shri Kulkarni
14:00 – 14:30  [Transient/Multi-messenger Science with Rubin Observatory (LSST)](#) – Fed Bianco
14:30 – 15:00  Transient/Multi-messenger Science with Other Optical Surveys – Paul Groot
15:00 – 15:30  Discussion and Questions
15:30 – 16:00  Break

**Session 8: Coordination and Co-Observing    Chair: Patricia Whitelock**

16:00 – 17:00  Bottlenecks IV: Coordination and Co-Observing
Guided Discussion Leaders: Robert Braun, Lisa Storrie-Lombardi
Topics:
[SKA: Coordinating across the EM Spectrum](#) (Robert Braun)
[LCO and the Role of Small Telescope Networks](#) (Lisa Storrie-Lombardi)
[The MASTER network](#) (Evgeny Gorbovskoy)
[J-GEM and OISTER](#) (Masayuki Yamanaka)
17:00 – 17:30  Summary Recommendations – Brad Cenko and Patricia Whitelock

**Wednesday, 5 February**

**Session 9: Compact Binary Mergers I  Chair: Mansi Kaliswal**

9:00 –  9:30    [Current and Future Ground-Based GW Detectors](#) – Laura Cadonati (presented by Johannes van den Brand)
9:30 – 10:00    [Theory of synchrotron radio afterglows](#) – Re'em Sari
10:00 – 10:30  Questions and Discussion
10:30 – 11:00  Break

**Session 10: Compact Binary Mergers II    Chair: Johannes van den Brand**

11:00 – 11:20  [Prompt Counterparts to GW Detections](#) – Binbin Zhang (via Zoom)
11:20 – 11:40  UV/Optical/IR Counterparts to GW Detections – Mansi Kasliwal

11:40 – 12:00  [X-ray/Radio Counterparts to GW Detections](#) – Tara Murphy
12:00 – 12:30  Summary Recommendations – Mansi Kaliswal and Johannes van den Brand
12:30 – 13:30  Lunch

**Session 11: Broadening the Scope of International Coordination Chair: Mansi Kasliwal**
13:30 – 13:50  [Transient/Multi-Messenger Opportunities in South Africa](#) – David Buckley
13:50 – 14:10  [Transient/Multi-Messenger Opportunities in Africa](#) – Jamal Mimouni
14:10 – 15:10  Bottlenecks V: International Funding
   Guided Discussion Leaders: John O'Meara and Roger Davies
   Topics:
   [Kavli Meeting on Ground/Space Coordination](#) (John O'Meara)
   [The IAU's role in International Cooperation](#) (Ewine van Dishoeck)
   [International Participation in CTA](#) (Ulisses Barres de Almeida)
15:10 – 15:30  NASA EM-GW Task Force Findings
   – Mansi Kasliwal, Valerie Connaughton and Brad Cenko
15:30 – 16:00  Summary Recommendations – Mansi Kasliwal

16:30 Wine Tasting followed by Workshop Dinner at [Durbanville Hills Winery](#)

**Thursday, 6 February**
**Session 12: Supermassive Black Holes    Chair: Samaya Nissanke**
 9:00 –  9:30   [Supermassive Black Hole Binaries with LISA](#) – Elena Rossi
 9:30 – 10:00  [Supermassive Black Hole Binaries with PTAs](#) – Ben Stappers
10:00 – 10:30  Discussion
10:30 – 11:00  Break

**Session 13: ToO Implementation    Chair: Samaya Nissanke**
11:00 – 12:00  Bottlenecks VI: ToO Implementation
   Guided Discussion Leaders: Pietro Ubertini and John Blakeslee
   Topics:
   [ToO Implementation on INTEGRAL](#) (Pietro Ubertini)
   [ToOs with ALMA](#) (John Carpenter)
   [ToOs with Rubin Observatory (LSST)](#) (Federica Bianco)
12:00 – 12:30 Summary Recommendations – Samaya Nissanke
12:30 – 13:30 Lunch

**Session 14: Theory and Machine Learning    Chair: Michelle Lochner**
13:30 – 14:00  [Simulations of SMBH Mergers](#) – Manuela Campanelli
14:00 – 14:30  Fundamental Physics with GW Detections – Samaya Nissanke
14:30 – 15:00  Discussion and Questions
15:00 – 15:30  Break

**Session 15: Computation    Chair: David Buckley**
15:30 – 15:50  [Machine Learning and Anomaly Detection for Transients](#) – Michelle Lochner
15:50 – 16:05  GCN – Scott Barthelmy
16:05 – 16:25  [Role of the IVOA](#) – Ada Nebot
16:25 – 17:15  Bottlenecks VII: Computation

          Guided Discussion Leaders: Patrick Brady, Ben Stappers
          Topics:
          SciMMA as a Center for MMA (Patrick Brady)
          The future of scientific computing (Manuela Campanelli)
17:15 – 17:30 Summary Recommendations – Michelle Lochner and David Buckley

**Friday, 7 February**
**Session 16: High-Energy Surveys    Chair: Valerie Connaughton**
 9:00 –  9:20   Current and Future High-Energy Facilities in India – Varun Bhalerao
 9:20 –  9:40   Current and Future High-Energy Facilities in China – Shuang-Nan Zhang
 9:40 – 10:30   Facilitated Discussion – Coordination in the High-Energy Sky
                 Panelists: Brad Cenko, Liz Hays, Pietro Ubertini
10:30 – 10:45  Summary Recommendations – Valerie Connaughton
10:45 – 11:15  Break
**Session 17: Diversity and workforce development    Chair: Roger Davies**
11:15 – 12:45  Bottlenecks VIII: Diversity and Workforce Development
               Guided Discussion Leaders: Liz Hays and Ranpal Gill
               Short talks followed by discussion:
               Present and Future Plans for Astronomy in Egypt (Somaya Saad)
               Diversity/Inclusion across a multi-national collaboration (Patrick Brady)
               Workforce development with IceCube (Ignacio Taboada)
               Lessons in Diversity/Inclusion from South Africa in preparation for SKA
               (Kavilan Moodley)
               Planning for the future astronomy workforce with Rubin Observatory (LSST)
               (Ranpal Gill)
               Breakout groups to discuss the following:
               A. How do scientists and engineers from smaller/less developed nations get involved in these big projects?
               B. How do we foster a diverse and inclusive environment in our international collaborations?
               C. How can we share best practice for improving diversity and inclusion? What training is needed? Can it be coordinated internationally?
               D. How do you measure the impact of programmes to improve diversity? How can it be done better?
12:45 – 13:00  Summary Recommendations – Roger Davies
13:00 – 13:45  Lunch
13:45 – 15:00  Wrap-up and Actions – Brad Cenko and Patricia Whitelock

## 2. Local Organizing Committee

Valencia Cloete (SAAO)

Daniel Cunnama (SAAO)

Christian Hettlage (SAAO)

Sivuyile Manxoyi (SAAO)
Nazli Mohamed (SAAO)
Glenda Snowball (SAAO)
Patricia Whitelock (chair; SAAO, UCT)

## 3. Organizations Represented

- [African Astronomical Society (AfAS)](#)
- [AstroSat](#)
- [Atacama Large Millimetre Array (ALMA)](#)
- [Cherenkov Telescope Array (CTA)](#)
- [Chinese Space Projects:](#)
    - Insight – Hard X-ray Modulation Telescope (HXMT)
    - Gravitational wave high-energy Electromagnetic Counterpart All-sky Monitor (GECAM)
    - Space Variable Objects Monitor (SVOM)
    - Einstein Probe (EP)
    - enhanced X-ray Timing and Polarimetry (eXTP)
- [Egyptian Large Optical Telescope (ELOT)](#)
- [European Southern Observatory (ESO)](#)
- [Fermi Gamma-ray Space Telescope](#)
- [Gamma-ray Coordinates Network (GCN)](#)
- [Gemini Observatory](#)
- [GROWTH India](#)
- [IceCube Neutrino Observatory](#)
- [International Astronomical Union (IAU)](#)
- [INTErnational Gamma-Ray Astrophysics Laboratory (INTEGRAL)](#)
- [International Virtual Observatory Alliance (IVOA)](#)
- [Japanese collaboration of Gravitational wave Electro-Magnetic follow-up (J-GEM)](#)
- [Las Cumbres Observatory (LCO)](#)
- [Laser Interferometer Gravitational-Wave Observatory (LIGO/Virgo)](#)
- [Major Atmospheric Gamma Imaging Cherenkov Telescopes (MAGIC)](#)
- [Mobile Astronomical System of TElescope Robots (MASTER)](#)
- [National Aeronautics and Space Administration (NASA)](#) – Headquarters
- [Neil Gehrels Swift Observatory (Swift)](#)
- [Optical and Infrared Synergetic Telescopes for Education and Research (OISTER)](#)

- [Sitian (司天) Project](#)
- [South African Astronomical Observatory (SAAO)](#)
- [Southern African Large Telescope (SALT)](#)
- [Square Kilometre Array (SKA)](#)
- [Transient Name Server (TNS)](#)
- [Vera C. Rubin Observatory (LSST)](#)
- [W. M. Keck Observatory](#)
- [Zwicky Transient Facility (ZTF)](#)

## 4. Participants

| | |
|---|---|
| Ulisses Barres de Almeida | Brazil |
| Scott Barthelmy | USA |
| Varun Bhalerao | India |
| John Blakeslee | USA |
| Patrick Brady | USA |
| Federica Bianco | USA |
| Robert Braun | UK |
| David Buckley | South Africa |
| Laura Cadonati | USA |
| Manuella Campanelli | USA |
| John Carpenter | Chile |
| Alberto Castro-Tirado | Spain |
| Brad Cenko | USA |
| Philip Charles | UK |
| Valerie Connaughton | USA |
| Roger Davies | UK |
| Rob Fender | UK |
| Anna Franckowiak | Germany |
| Avishay Gal-yam | Israel |
| Ranpal Gill | USA |
| Evgeny Gorbovskoy | Russia |
| Paul Groot | Netherlands |
| Liz Hays | USA |
| Rob Ivison | Germany |
| Mansi Kasliwal | USA |
| Shri Kulkarni | USA |
| Michelle Lochner | South Africa |
| Christopher Martin | USA |
| Jamal Mimouni | Algeria |
| Kavilan Moodley | South Africa |
| Tara Murphy | Australia |
| Ada Nebot | France |
| Samaya Nissanke | Netherlands |
| John O'Meara | USA |
| Stephen Potter | South Africa |
| Elena Rossi | Netherlands |
| Somaya Saad | Egypt |
| Re'em Sari | Israel |
| Roberto Soria | China |
| Sarah Burke Spolaor | USA |
| Ben Stappers | UK |
| Lisa Storrie-Lombardi | USA |
| Ignacio Taboada | USA |
| Pietro Ubertini | Italy |
| Johannes van den Brand | Netherlands |
| Ewine van Dishoeck | Netherlands |
| Patricia Whitelock | South Africa |
| Patrick Woudt | South Africa |
| Masayuki Yamanaka | Japan |
| Binbin Zhang | China |
| Shuang-Nan Zhang | China |

## 5. Acronyms used in this report

| | |
|---|---|
| AEON | Astronomical Event Observatory Network |
| ALMA | Atacama Large Millimeter/submillimeter Array |
| ASAS-SN | All Sky Automated Survey for SuperNovae |
| ASKAP | Australian Square Kilometre Array Pathfinder |
| ATLAS | Asteroid Terrestrial-impact Last Alert System |
| AUI | Associated Universities, Inc. |
| AURA | Association of Universities for Research in Astronomy |
| CDS | Strasbourg astronomical Data Centre |
| CSDC | Community Science and Data Center |
| CTA | Cherenkov Telescope Array |
| eROSITA | extended ROentgen Survey with an Imaging Telescope Array |
| EPOS | European Plate Observing System |
| ERIC | European Research Infrastructure Consortium |
| ESCAPE | European Science Cluster of Astronomy & Particle physics ESFRI |
| ESFRI | European Strategy Forum on Research Infrastructures |
| ESO | European Southern Observatory |
| ECWG | IAU Executive Committee's Working Group |
| GCN | Gamma-ray Coordinates Network |
| GRANDMA | Global Rapid Advanced Network Devoted to Multi-messenger Addicts |
| GROWTH | Global Relay of Observatories Watching Transients Happen |
| HST | Hubble Space Telescope |
| HAWCO | High-Altitude Water Cherenkov Observatory |
| IAU | International Astronomical Union |
| IGO | Inter-Governmental Organization |
| INTEGRAL | INTErnational Gamma-Ray Astrophysics Laboratory |
| ISYA | International School for Young Astronomers |
| IVOA | International Virtual Observatory Alliance |
| LEO | Low Earth Orbit |
| LHC | Large Hadron Collider |
| LIGO | Laser Interferometer Gravitational-Wave Observatory |
| LISA | Laser Interferometer Space Antenna |
| LOC | Local Organizing Committee |
| LSST | Legacy Survey of Space and Time |
| MDN | Multi-messenger Diversity Network |
| MeerKAT | Meer Karoo Array Telescope |
| NASA | National Aeronautics and Space Administration |
| NED | NASA/IPAC Extragalactic Database |
| NOIRLab | National Optical-Infrared Astronomy Research Laboratory |
| NuSTAR | Nuclear Spectroscopic Telescope Array |
| OAD | IAU Office of Astronomy for Development |
| OYA | IAU Office for Young Astronomers |
| Pan-STARRS | Panoramic Survey Telescope and Rapid Response System |
| PIRE | Partnerships for International Research and Education |
| SAAO | South African Astronomical Observatory |
| SARAO | South African Radio Astronomy Observatory |
| SCiMMA | Scalable Cyberinfrastructure for Multi-Messenger Astrophysics |
| SED | Spectral Energy Distribution |
| SKA | Square Kilometre Array |
| SOAR | Southern Astrophysical Research Telescope |
| SOC | Scientific Organizing Committee |
| SRG | Spectrum Roentgen Gamma |
| TACH | [Time-Domain Astronomy Coordinate Hub](#) |
| TNS | [Transient Name Server](#) |
| ToO | Target of Opportunity |
| UK | United Kingdom |
| USA | United States of America |
| VLA | [Very Large Array](#) |
| VLBI | Very Long Baseline Interferometer |
| XMM | [X-ray Multi-mirror Mission](#) |
| ZTF | [Zwicky Transient Facility](#) |